\documentclass{article}
\usepackage{amsmath}
\usepackage{latexsym}

\begin{document}

\title{Cosmological energy in a thermo-horizon and the first law}
\author{C.\ Barbachoux\thanks{%
barba@ccr.jussieu.fr}, J.\ Gariel and G.\ Le Denmat \\
LERMA, UMR CNRS 8112\\
Universit\'{e} P. et M. Curie ERGA, B.C. 142\\
3, Rue Galil\'{e}e, 94 200 Ivry, France}
\maketitle

\begin{abstract}
We consider a cosmological horizon, named thermo-horizon, to which
are associated a temperature and an entropy of Bekenstein-Hawking
and which obeys the first law for an energy flow calculated through
the corresponding limit surface. We point out a contradiction
between the first law and the definition of the total energy
contained inside the horizon. This contradiction is removed when the
first law is replaced by a Gibbs' equation for a vacuum-like
component associated to the event horizon.
\end{abstract}

\section{Introduction}

The generalization of the Thermodynamics of black holes (BH) to cosmological
horizons represents an important stake to understand different issues in
cosmology such as the nature of the dark energy (DE) in relation with the
problems of the cosmological constant (CC) and of the vacuum energy, the
acceleration of the present universe, the coincidence problem and the early
inflation.

This generalization was first introduced for de Sitter spacetime \cite%
{article1}. Thereafter, it was tentatively extended to quasi-de Sitter FRW
spacetimes in different frameworks (see \cite{article2}- \cite{article5}).
In an interesting approach, Bousso \cite{article6,article7} considers the
flow of energy through the horizon as a null surface. He interprets the
variation of the entropy of the horizon through the variation of its surface
as the response of the horizon to the flux of energy, in the same way as the
\textquotedblleft first law" of the BH.

Following this approach, several authors (see for example \cite%
{article8,article9}) have estimated that the apparent horizon (a.h.) is the
only limit surface (excluding other horizons such as the event horizon
(e.h.)) having coherent thermodynamical properties to address problems such
as the nature of the DE.

Our main goal is to shed some light on the contradiction between the amount
of energy calculated from the first law as defined in \cite%
{article6,article7} and the definition of the energy contained inside the
horizon, independently of the choice of the thermo-horizon (t.h.).

We restrict our study to a spatially flat FRW spacetime, which is the
starting point of other studies (non spatially flat spacetimes, cases with
interactions,...).

After a brief review of the definition of a t.h. in a $Q$-space introduced
in \cite{article6,article7}, we show that any t.h. obeys the second law
(Section 2). In section 3, we present the contradiction between the amount
of energy derived from the first law and the definition of the energy inside
the horizon. We then show that this contradiction is resolved in a
thermodynamical model for a DE \cite{article4,article4bis} based on the e.h.
(Section 4).

\section{Definition of a thermo-horizon}

In a spatially flat FRW spacetime
\begin{equation}
ds^{2}=-dt^{2}+a(t)^{2}(dr^{2}+r^{2} d\Omega^{2}),
\end{equation}
the dynamical evolution of the scale factor $a(t)$ is given for a perfect
fluid with energy density $\rho $ and pressure $P$ by
\begin{eqnarray}
\left(\frac {\dot a }{a}\right)^2=H^2&=&\chi\frac {\rho}{3},  \label{eq0-1}
\\
\frac {\ddot a}{a}&=&-\frac {\chi}{6} (\rho+3P) ,  \label{eq0-2}
\end{eqnarray}
where $\chi =8\pi $ is the Einstein constant, with $G=1$ and $c=1$. The
equation of state (EoS) $\omega $ of the fluid is given by $P=\omega \rho $
and we introduce the parameter $\varepsilon =\frac{3}{2}(1+\omega )$. In the
following, we restrict our study to the $Q$-space \cite{article7}, namely
accelerated universes, for which $0<\varepsilon <1$. Extending the reasoning
of \cite{article7}, we consider an horizon (null surface) with a given
radius $L$.\ According to the first law, the flow of energy through this
surface is given by
\begin{equation}
-\overset{\cdot }{E}=4\pi L^{2}\rho (1+\omega )=T\overset{\cdot }{S}.
\label{eq1}
\end{equation}

We assume that we can associate a temperature $T$ and an entropy $S$ to the
dynamical horizon of radius $L$, given by the relations of
Bekenstein-Hawking for a BH or a de Sitter horizon
\begin{equation}
T=\frac{1}{2\pi L}\text{ , \ and \ \ }S=\pi L^{2}.  \label{eq2}
\end{equation}%
Any horizon of radius $L$ with a temperature and a entropy given by (\ref%
{eq2}) and which obeys the first law (\ref{eq1}) is called thermo-horizon
(t.h.).

Using (\ref{eq2}), we obtain directly $T\overset{\cdot }{S}=\overset{\cdot }{%
L}$ and Eq. (\ref{eq1}) becomes
\begin{equation}
\varepsilon L^{2}H^{2}=\overset{\cdot }{L}.  \label{eq3}
\end{equation}%
With our notations, the Eq. (\ref{eq0-2}) is given by
\begin{equation}
\overset{\cdot }{\left( \frac{1}{H}\right) }=\varepsilon ,  \label{eq4}
\end{equation}%
and the first law (\ref{eq3}) rewrites
\begin{equation}
\overset{\cdot }{H}\text{ }=\overset{\cdot }{(\frac{1}{L})}.  \label{eq5}
\end{equation}%
After integration, this equation leads to
\begin{equation}
HL-1=CL,  \label{eq6}
\end{equation}%
where $C$ is a constant. Eq. (\ref{eq6}) establishes a general relation
between the t.h. $L$ and the a.h. $R_{A}=\displaystyle\frac{1}{H}$ which is
satisfied by any thermo-horizon of radius $L$ without restriction on $%
\varepsilon $ (in particular without assuming $\varepsilon =constant$). With
the constant $C$, this relation is more general than the Eq. (28) of \cite%
{article7}.

If $L$ is the a.h., then $L=\displaystyle\frac{1}{H}=R_{A}$,
implying $C=0$. Conversely, only $C=0$ leads to
$L=\displaystyle\frac{1}{H}$. Therefore the a.h. obeys the first law
(\ref{eq1}) if and only if $C=0$. This special case only is
considered by \cite{article7}.

More generally, any horizon $L$ defined by (\ref{eq6}) with a temperature
and an entropy given by (\ref{eq2}) is a t.h. and it obeys the first law (%
\ref{eq1}).

The equation (\ref{eq3}) can be rewritten with the help of (\ref{eq6})
\begin{equation}
\overset{\cdot }{L}=\varepsilon (1+CL)^{2}.  \label{eq7}
\end{equation}%
Any t.h. verifies this equation. Using Eq. (\ref{eq7}), $L$ is strictly
increasing in the $Q$-space (accelerated universe) where $0<\varepsilon <1$.
The same result can be derived for the entropy $S$ given by (\ref{eq2}).

\section{Energy in a thermo-horizon and the first law}

On one side, the total amount of energy contained inside the a.h. for a
spatially flat FW space-time is (e.g. \cite{article8} before Eq. (20))
\begin{equation}
E=\rho \frac{4\pi }{3}R_{A}^{3}=\frac{R_{A}}{2}.  \label{eq3-1}
\end{equation}%
Let us remark that in \cite{article9} this relation is used for a
non-spatially flat FRW spacetime, albeit no more valid in this case.

On the other side, using (\ref{eq1}) and (\ref{eq2}), the first law applied
to the a.h. considered as a t.h. leads to
\begin{equation}
-\overset{\cdot }{E}\text{ }=\overset{\cdot }{R_{A}},  \label{eq3-2}
\end{equation}%
where $-\overset{\cdot }{E}$ is the total amount of energy crossing the a.h.
by unit of time. According to the conservation of the energy, this amount of
energy is equal to the variation of the total energy (\ref{eq3-1}) per unit
time, $\overset{\cdot }{E}=\displaystyle\frac{\overset{\cdot }{R_{A}}}{2}$.
This result is in contradiction with (\ref{eq3-2}) except when $R_{A}$ is
constant, which corresponds to a de Sitter spacetime where the a.h.
identifies with the e.h..

This result is not restricted to the a.h. and can be extended to any t.h..
Using (\ref{eq4}), the left hand side of (\ref{eq1}) becomes for a t.h. of
radius $L$
\begin{equation}
-\overset{\cdot }{E}=L^{2}H^{2}\varepsilon =-L^{2}\overset{\cdot }{H},
\label{eq3-3}
\end{equation}%
while the total energy inside the horizon is
\begin{equation}
E=\frac{1}{2}H^{2}L^{3}.  \label{eq3-4}
\end{equation}%
Differentiating (\ref{eq3-4}) and equating with (\ref{eq3-3}), we obtain
with (\ref{eq5})
\begin{equation}
\frac{3}{2}(HL)^{2}-HL+1=0,  \label{eq3-5}
\end{equation}%
where $\overset{\cdot }{H}$ $\neq 0$ has been assumed. No real root can be
found for this equation. In particular, $L=\displaystyle\frac{1}{H}$ is not
a solution. Therefore, the above contradiction can only be removed for $%
\overset{\cdot }{H}=0$, namely for a de Sitter spacetime.

\section{Thermodynamical model of the event horizon for the dark energy}

The preceding results are independent of the underlying model for the DE.
They depend only on the assumptions of the existence of a temperature and an
entropy associated to a t.h. through the relation (\ref{eq2}) and of the
validity of the extension of the first law of BHs (\ref{eq1}) to
cosmological t.hs.. With these assumptions, we obtain (\ref{eq6}) (assuming $%
C=0$), as demonstrated by \cite{article6,article7} and by \cite{article8},
in sections II-A and II-B.

In section II-B of \cite{article8}, the authors consider only the specific
model for the DE developed in \cite{article5}. Let us emphasize that their
results can be obtained independently of any model for the DE (see Section
2) because the demonstration involves only the density of the total energy $%
\rho $. Consequently, the reasoning developed in \cite{article8} cannot
question the validity of the model assumed for the component DE and in
particular the approach proposed in \cite{article5}. The first law (\ref{eq1}%
) is a relation between the density of energy $\rho $ and the entropy. It
does not involve the density of energy of the DE $\rho _{\Lambda }$ and
therefore cannot be used to discuss or refute its expression.

In \cite{article8}, the authors emphasize the apparent discrepancy between
the horizon chosen as IR cut-off in the expression of $\rho _{\Lambda }$ in
the holographic model of Li \cite{article5}, which is the e.h. $r$
\begin{equation}
\chi \rho _{\Lambda }=\displaystyle\frac{3c^{2}}{r^{2}}\text{ ,}
\label{eq4-1}
\end{equation}%
and the t.h. $L$ which must be the a.h. $R_{A}$ in order to satisfy the
first law. They suggest that the e.h. $r$ should be identical to the t.h. $L$
obeying the first law. So, they implicitly assume that in any holographic
model for $\rho _{\Lambda }$, i.e. $\chi \rho _{\Lambda }=3/l^{2}$, we have
to choose for cut-off $l$ the same horizon as the t.h. $L$ obeying the first
law.

Let us discuss the full consequences of the assumption $l=L$. On one hand,
by setting $L=l=r$ in the expression of the flux of energy in the first law (%
\ref{eq1}), the first law is no more satisfied (see Section 3). On the other
hand, if we choose $L=l=R_{A}$ , we obtain for the holographic model:
\begin{equation}
\chi \rho _{\Lambda }=\frac{3}{R_{A}^{2}}=3H^{2}.  \label{eq4-2}
\end{equation}%
which with (\ref{eq0-1}) implies $\rho =\rho _{\Lambda }$, excluding any
other contribution to the total energy (dark matter (DM), dust,
radiation,...) in contradiction with the present observations where the DM
takes a non negligible part (about $1/3$) of the energy of the universe.

Two points of view can be followed to solve this dilemma:

\begin{itemize}
\item[i)] First, we can choose to preserve the first law (\ref{eq1}) and
propose an holographic model of the form (\ref{eq4-2}) albeit not compatible
with the observations \cite{article5,article11}.

\item[ii)] Secondly, we can consider the holographic model of the DE (\ref%
{eq4-1}) compatible with the present observations (acceleration and EoS
today $\omega _{\Lambda }\simeq -1$ \cite{article4bis,article5,article10})
and modify the first law (\ref{eq1}) to be compatible with the chosen model
for the DE.
\end{itemize}

Because compatible with the observational features, the second alternative
is more reasonable . This leads naturally to question the first law (\ref%
{eq1}) which seems to fail because, as seen in Section 3, it contradicts the
definition (\ref{eq3-1}) of the total energy in the horizon. The first law
must be modified in order to include the DE through a model linking the DE
with the chosen horizon. To achieve this goal, we propose a Gibbs' equation
describing the thermodynamics of the component DE instead of the first law (%
\ref{eq1}). This approach was introduced in the model \cite%
{article4,article4bis}, where a DE component with an energy density of type (%
\ref{eq4-1}) (with $c=1$) and an EoS of vacuum-type were considered
\begin{equation}
\frac{\Lambda }{\chi }=\rho _{\Lambda }=-P_{\Lambda }=3/\chi r^{2},
\label{eq4-3}
\end{equation}%
with $r$ the radius of the e.h.. In this model, the Gibbs' equation relative
to this component DE is given at the specific level by
\begin{equation}
T_{\Lambda }ds_{\Lambda }=d\varepsilon _{\Lambda }+P_{\Lambda }dv_{\Lambda },
\label{eq4-3bis}
\end{equation}%
where $\varepsilon _{\Lambda }\equiv \displaystyle\frac{\Lambda }{n_{\Lambda
}\chi }$ is the specific energy, $v_{\Lambda }\equiv \displaystyle\frac{1}{%
n_{\Lambda }}$ the specific volume and $n_{\Lambda }$ the number density.
With the EoS (\ref{eq4-3}), the Eq. (\ref{eq4-3bis}) becomes
\begin{equation}
\frac{\overset{\cdot }{\Lambda }}{n}=\chi T_{\Lambda }\overset{\cdot }{%
S_{\Lambda }}\text{ }=-\overset{\cdot }{r},  \label{eq4-4}
\end{equation}%
with $T_{\Lambda }=\displaystyle\frac{1}{2\pi r}$. After integration, we
obtain for the specific entropy
\begin{equation}
\text{ }S_{\Lambda }=-\pi r^{2}+K,  \label{eq4-5}
\end{equation}%
where $K$ is a constant. In \cite{article4,article4bis} the local equation
of the conservation of the energy $u_{\beta }\nabla _{\alpha }T^{\alpha
\beta }=0$ is used instead of the expression of the global energy $E$ and
without assuming any peculiar expression for the entropy.

Considering the global energy of the DE component inside the horizon as in
\cite{article8,article9}, $E_{\Lambda }=\rho _{\Lambda }\displaystyle\frac{%
4\pi }{3}r^{3}=\displaystyle\frac{r}{2}$, the Gibbs'equation associated to
this component becomes
\begin{equation}
dE_{\Lambda }=T_{\Lambda }dS_{\Lambda }-P_{\Lambda }dV_{\Lambda }\text{ , or
}\overset{\cdot }{E_{\Lambda }}=T_{\Lambda }\overset{\cdot }{S_{\Lambda }}%
-P_{\Lambda }\overset{\cdot }{V_{\Lambda }}.  \label{eq4-6}
\end{equation}%
With (\ref{eq4-3}) and (\ref{eq4-5}), this Eq. leads to $\overset{\cdot }{%
E_{\Lambda }}=\displaystyle\frac{\overset{\cdot }{r}}{2}$, in full
agreement with the previous definition of $E_{\Lambda }=\displaystyle\frac{%
r}{2}$.

The previous results show that the contradiction described in Section 3 is
essentially related to the mainly questionable assumption of the validity of
the expression of the static entropy $S=\pi L^{2}$ of the BH for the
cosmological horizon \cite{article6,article7}. On the contrary, our
expression (\ref{eq4-5}) of $S_{\Lambda }$ is not postulated but deduced
from the Gibbs' equation for the model (\ref{eq4-3}) of the DE \cite%
{article4,article4bis}, which is supported by the holographic approach \cite%
{article5}.

By comparison of the two Eqs. (\ref{eq2}) and (\ref{eq4-5}), let us first
note the difference of sign in the expression of the entropy. This
difference is in accordance with the predictions because in the first case,
the observer is outside the limit-surface of the BH and looses information,
while in the second case (cosmological case), the observer is inside the
limit-surface. This difference of sign is pointed out in \cite%
{article6,article7} where it is attributed to the exchange of energy $-dE$
rather than to $TdS$, while we interpret it as a lost (or a gain) of
information from the limit-surface.

Secondly, let us remark the presence of a pressure term in the
Gibbs' equation (\ref{eq4-6}) albeit not appearing in the first law
(\ref{eq1}). This term takes into account the fact that the DE has a
(negative) pressure, and that the transformation is not isochore
because the surface varies. This term is
 responsible for
the sign ``minus" in the expression of the entropy Eq. (\ref%
{eq4-5}). This point relates the two preceding remarks and strengthens the
consistency of this approach. The existence of such a term was recently
considered and discussed in another context for the first law of the BH by
\cite{article12}.

\section{Conclusion}

By a general demonstration independent of the underlying model for the DE,
we show in this article a contradiction (except in de Sitter) between the
first law introduced in \cite{article6,article7} for the thermo-horizon that
leads to $\overset{\cdot }{-E}$ $=\overset{\cdot }{R_{A}}$, and the
definition of the total energy in the horizon $E=\displaystyle\frac{R_{A}}{2}
$ . To solve this contradiction, we propose to replace the first law by a
Gibbs' equation for the DE component, which is naturally associated to the
e.h. in the model \cite{article4,article4bis}, later supported by an
holographic model \cite{article5} in an independent approach.

\end{document}